\documentclass[manuscript]{acmart}
\AtBeginDocument{%
  \providecommand\BibTeX{{%
    \normalfont B\kern-0.5em{\scshape i\kern-0.25em b}\kern-0.8em\TeX}}}

\setcopyright{acmcopyright}
\copyrightyear{2023}
\acmYear{2023}
\acmDOI{XXXXXXX.XXXXXXX}

\acmConference[dg.o '23]{Annual International Conference on Digital Government Research}{July 11--14,
  2023}{Gdańsk, Poland}
\acmBooktitle{dg.o '23: Annual International Conference on Digital Government Research, July 11--14, 2023, Gdańsk, Poland} 
\acmPrice{0.00}
\acmISBN{}

\usepackage{multirow}
\usepackage{caption}
\usepackage{hyperref}
\usepackage{subcaption}

\begin{document}

\title[Improving City Life via Data-driven Approach]{Improving City Life via Legitimate and Participatory Policy-making: A Data-driven Approach in Switzerland}

\author{Thomas Wellings$^{1}$}
\author{Srijoni Majumdar$^{1}$}
\author{Regula Hänggli Fricker$^2$}
\author{\\Evangelos Pournaras$^{1}$}

\affiliation{%
  \institution{\\$^1$ University of Leeds}
  \country{United Kingdom}
}
%\email{\{t.wellings,s.majumdar,e.pournaras\}@leeds.ac.uk}
\affiliation{%
  \institution{$^2$ University of Fribourg}
  \country{Switzerland}
}


%\email{webmaster@marysville-ohio.com}
% \affiliation{\newline
%   \institution{$^1$ School of Computing, University of Leeds, UK} \newline
%     \institution{$^2$  Department of Mass Media and Communication Research, University of Fribourg, Switzerland}}
\renewcommand{\shortauthors}{Wellings et al.}

\begin{abstract}
This paper introduces a novel data-driven approach to address challenges faced by city policymakers concerning the distribution of public funds. Providing budgeting processes for improving quality of life based on objective (data-driven) evidence has been so far a missing element in policy-making. This paper focuses on a case study of 1,204 citizens in the city of Aarau, Switzerland, and analyzes survey data containing insightful indicators that can impact the legitimacy of decision-making. Our approach is twofold. On the one hand, we aim to optimize the legitimacy of policymakers' decisions by identifying the level of investment in neighborhoods and projects that offer the greatest return in legitimacy. To do so, we introduce a new context-independent legitimacy metric for policymakers. This metric allows us to distinguish decisive vs. indecisive collective preferences for neighborhoods or projects on which to invest, enabling policymakers to prioritize impactful bottom-up consultations and participatory initiatives (e.g., participatory budgeting). The metric also allows policymakers to identify the optimal number of investments in various project sectors and neighborhoods (in terms of legitimacy gain). On the other hand, we aim to offer guidance to policymakers concerning which satisfaction and participation factors influence citizens' quality of life through an accurate classification model and an evaluation of relocations. By doing so, policymakers may be able to further refine their strategy, making targeted investments with significant benefits to citizens' quality of life. These findings are expected to provide transformative insights for practicing direct democracy in Switzerland and a blueprint for policy-making to adopt worldwide.

%Findings from a classification model are used to explore citizens' quality of life, which is based upon citizens satisfaction and participation data. In addition, the legitimacy indicators outlined within this paper may contribute to an informed decision making process, offering guidance of when to prioritize participatory interventions (such as participatory budgeting. The paper should contribute to understandings of legitimacy and the steps that policymakers can take to optimize legitimacy within the local budgeting process. 

\end{abstract}

\begin{CCSXML}
\begin{CCSXML}
<ccs2012>
   <concept>
       <concept_id>10010147</concept_id>
       <concept_desc>Computing methodologies</concept_desc>
       <concept_significance>500</concept_significance>
       </concept>
   <concept>
       <concept_id>10010405.10010455.10010460</concept_id>
       <concept_desc>Applied computing~Economics</concept_desc>
       <concept_significance>300</concept_significance>
       </concept> 
   <concept>
       <concept_id>10002944.10011123.10010912</concept_id>
       <concept_desc>General and reference~Empirical studies</concept_desc>
       <concept_significance>500</concept_significance>
       </concept>
   <concept>
       <concept_id>10003456.10003462</concept_id>
       <concept_desc>Social and professional topics~Computing / technology policy</concept_desc>
       <concept_significance>500</concept_significance>
       </concept>
 </ccs2012>
\end{CCSXML}

\ccsdesc[500]{Computing methodologies}
\ccsdesc[300]{Applied computing~Economics}
\ccsdesc[500]{General and reference~Empirical studies}
\ccsdesc[500]{Social and professional topics~Computing / technology policy}

\keywords{legitimacy, policy-making, decision-making, data science, participatory budgeting, quality of life, participation}



\maketitle

\section{Introduction}
\label{sec:intro}

In liberal democracies, political authority relies on public support. However, Saar~\cite{Saar2016} argues that policymakers often suffer from a legitimacy deficit and may seek to mitigate such a deficit through an increased focus on the wishes of the citizenry. To gauge the wishes of the citizenry, in relation to the distribution of a local budget, participatory budgeting has become a popular method. Participatory budgeting may be beneficial for policymakers as it can increase input legitimacy (the exercise of collective self-governance) ~\cite{schmidt2020,Hanggli2021human}. 

Many countries, including Poland, Spain, Brazil, Germany, and Switzerland, to name a few, conduct participatory budgeting to ask people to participate in a decision-making process, in which people propose project ideas and vote for the ones that they prefer to be implemented in a city or neighbourhood~\cite{sintomer2008,maeroe2021}. However, participatory budgeting is not always feasible or straightforward to apply. It is often a costly and time-consuming process, which has several challenges, such as lack of systematic applicability, limited support and expertise from city staff, and increased participation costs~\cite{pinnington2009}. In addition, without an optimised process, participatory budgeting can lead to citizen's disappointment and gradual abandonment - as has been observed in Brazil \cite{bezerra2022has}. Therefore, steps should be undertaken to determine the feasibility of undertaking participatory budgeting within a given locality and to maximize the expected legitimacy returns.

Traditionally, city authorities distribute funds in a top-down manner, determined on the basis of policy priorities~\cite{ljungman2009} or popularity/demand of the project(s). Top-down decision making comes with its own challenges. For example, when policy priorities differ to popular demands, policymakers may face a challenge to their output legitimacy (top-down authority may oppose what benefits common good~\cite{schmidt2020}). 

In this paper, we argue that policymakers can maximize their legitimacy through an informed, data-driven and evidence-based decision-making process, which may offer guidance of when to prioritize participatory interventions and how to optimize investment. To do so, we propose a novel context-independent measure of legitimacy. Policy-makers can utilize our approach to determine the optimal number of project sectors and neighborhoods that guarantees legitimacy improvement. Should funding of the optimal number of project sectors not be available, then a participatory intervention may be beneficial to prioritize. 

In addition, policymakers often face the issue of being unable to fulfill the funding demands of citizenry. In such circumstances, spending prioritization can be a challenge for policymakers. We address such an issue by providing policymakers guidance regarding the type of project sectors that have the greatest impact on citizens' quality of life. To do so, we present a classification model to identify and explain which participation and satisfaction (independent variables) lead to an improvement in citizens' overall quality of life (dependent variable), and conduct an analysis of which public services citizens place importance on when relocating. Our approach should provide the local authorities guidance regarding the allocation of public spending, offering a blueprint for legitimate targeted investments.

%%%%%%%%%%%%%%%%%%%%%%%%%%%%%%%%%%%%%%%%%%%%%%%%

To summarize, the new findings from this paper are invaluable for city authorities to prioritize participatory approaches (e.g. participatory budgeting) with the highest positive impact on legitimacy. These findings are extracted by analysing a high-quality survey from the city of Aarau in Switzerland with 1,204 participants using advanced machine learning methods.  

As a summary, this paper addresses the following research questions:
\begin{itemize}
    \item \textbf{RQ1:} How to prioritize bottom-up participatory interventions to improve legitimacy of policy-making? 
 
    \item \textbf{RQ2:} Which indicators of satisfaction and participation explain citizens' overall quality of life?
    
    \item \textbf{RQ3:} Are citizens' relocation to different neighbourhoods associated with quality of life improvements? 
    
\end{itemize}

The contributions of this paper are summarized as follows:

\begin{itemize}
    \item A context-independent metric to measure the legitimacy of allocating funds to different projects/sectors and neighborhoods. 
    \item An optimization heuristic to select a sufficient number of popular project sectors to fund so that legitimacy improves before saturation.
    \item A classification model, linked to legitimacy, that explains which satisfaction and participation factors improve the overall citizens' quality of life.
    \item Metrics for the improvement of quality of life as a result of citizens' neighbourhood relocations within a city. 
    \item A case study for the city of Aarau in Switzerland, whose findings provide a blueprint for designing and running a novel participatory budgeting campaign. 
    \item An open dataset~\cite{Wellings2023} based on the survey data collected and the analysis performed.
\end{itemize}

This paper is outlined as follows: Section~\ref{sec:related-work} reviews related work. Section~\ref{sec:Methodology} outlines the methodological approach. Section~\ref{sec:Results} illustrates the empirical results and findings. Section~\ref{sec:Conclusion} concludes this paper and outlines directions for future work.

\section{Related Literature Review}\label{sec:related-work}

In this section, we offer an overview of related literature, with a focus on quality of life and public budgeting. In doing so, we also consider the role of political legitimacy. Political legitimacy is a relatively ambiguous term~\cite{buchanan2002}, which falls into what Gaille outlined as an essentially contested concept~\cite{gaille1955}. In general, legitimacy can be understood as an assumption that the actions and decisions made are desirable, acceptable, or proper within a system which is socially connected by beliefs, norms, and values~\cite{suddaby2017}. The scope of the empirical measurements of this paper focus on normative measurements of legitimacy \cite{peter2010,andersen2012legitimacy,Hanggli2021human}. 

Quality of life has a demonstrated link with legitimacy for policymakers~\cite{bay1968needs, hanberger2003public}. However, Leanard~\cite{leonard2009} highlights how quality of life alone is not enough to lead to positive assessments of the local authority. Namely, even when citizens are satisfied with public services (that may be used as an indicator for quality of life), there is still the possibility to hold a negative attitude towards policymakers, bringing into question policymakers' legitimacy. It has been demonstrated that participatory interventions may aid in increasing the legitimacy of the local authority, the quality of life for citizens and satisfaction with the political processes~\cite{swaner2017trust}. In addition, participatory budgeting has a demonstrated link with a decrease in hostility from the public~\cite{magliacani2020public}. In this sense, \textit{both} quality of life metrics and the decision-making process require attention by policymakers. To determine quality of life, we use satisfaction and participation data, which allows for informed investment within particular sectors. Here, we offer a basis for our approach, considering previous literature. 

Deniz et al.~\cite{deniz2019} examined the relationship between quality of life, satisfaction with life and multidimensional perceived social support in people aged 65 years and older. Importantly, Deniz et al. demonstrate a moderate correlation between quality of life and satisfaction, which may offer support for our decision to explore the satisfaction of public services and quality of life within our paper. In addition, research has pointed to social factors, that we have also incorporated into our measurement (through participation data). For example, Medvedev et al.~\cite{medvedev2018} studied the relationship between happiness, subjective well-being, and quality of life, and found that social relationships and environmental factors impact quality of life. In addition, Macke et al.~\cite{macke2018s} conducted interviews of 400 residents of Curitiba, a city in Southern Brazil, to investigate the major elements behind people's satisfaction with their city. Notably, they find that socio-structural relations, environmental well-being, material well-being and community integration can have an impact on quality of life. 

\begin{figure}[!htb]
\centering
  \includegraphics[scale=0.45]{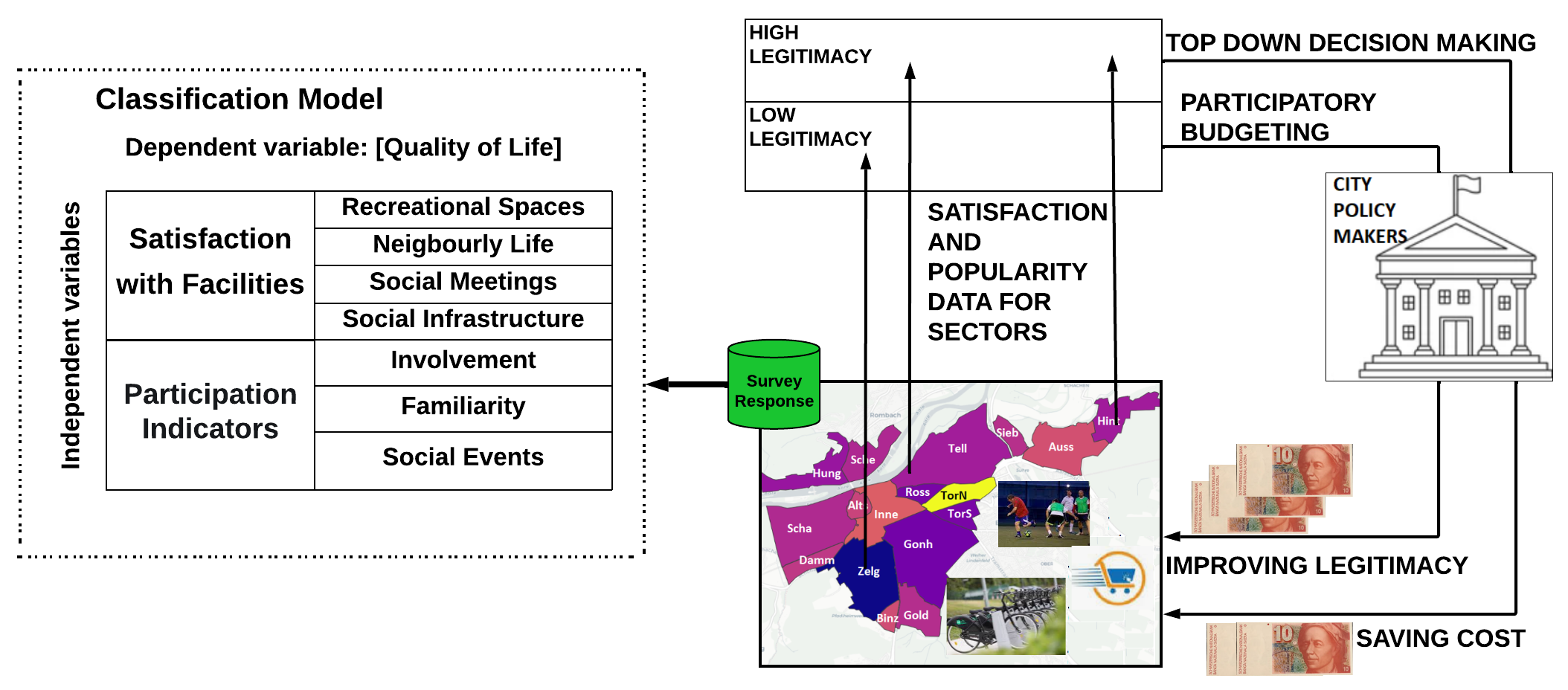}
   \caption{The proposed data-driven model to improve legitimacy in policy making.}
  \label{fig:archi}
\end{figure}

Our decision to explore relocation as a factor that may impact quality of life is supported by the findings of Morris et al.~\cite{morris2019} and Wong~\cite{wong2001}. Specifically, Morris et al.~\cite{morris2019} conducted a study on the emotional satisfaction by comparing the residence in principal city and suburbs, finding that geographical location may impact a citizen's quality of life. Moreover, Wong~\cite{wong2001} compared traditional quality of life measurements against location in relation to citizens perceptions of local economic development, finding that location was considered to be a significant factor. In our paper, we have explored intra-city relocation (the movement of citizens from one neighborhood of Aarau to another). In doing so, we are able to explore the improvement or deterioration of satisfaction with the sectors identified as an indicator of a citizen's quality of life.

\section{Methodology}
\label{sec:Methodology}
In this section, we first introduce the dataset, which has been made publicly available~\cite{Wellings2023}. Later on we present an approach to answer the research questions formulated in Section~\ref{sec:intro}. The proposed data-driven policy-making model for legitimacy is outlined in Fig.~\ref{fig:archi}.

\subsection{Dataset}

\begin{figure}[!htb]
  \includegraphics[width=0.48\textwidth]{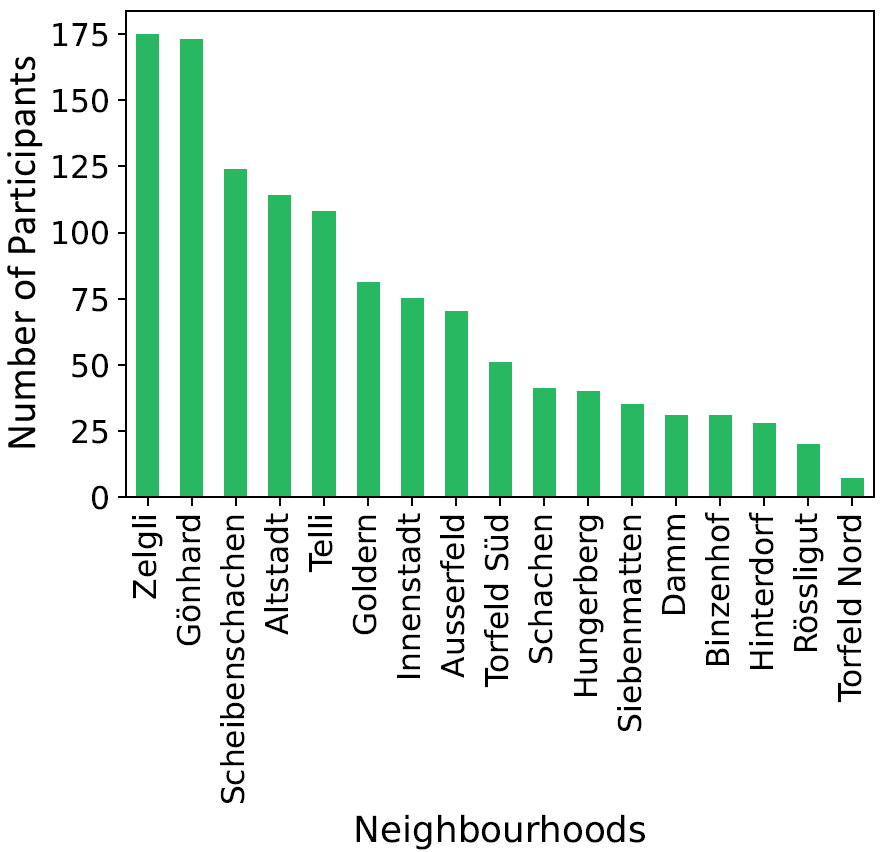}
\caption{Number of participants from each neighbourhood.}
\label{fig:resp_from_each_neighborhood}
\end{figure}

The dataset is derived from an online survey conducted in Aarau, Switzerland, on behalf of the local authority between 18 March and 24 April 2020, see~\cite{Wellings2023}. The survey was open to the 22,032 citizens of Aarau, and 1,204 usable responses were collected. The questionnaire featured a combination of open-ended and closed-ended (multiple-choice and Likert scale) questions. The survey included questions related to demographics, quality of life, neighborhoods, socializing/networking, and mobility. Additionally, participants were asked to propose project ideas they would like to see implemented in their neighborhoods (detailed information on the questions is provided in Table~\ref{tab:q_cat}). Representing 17 different neighborhoods in Aarau, the respondents comprised 548 males, 534 females, 8 individuals with unspecified gender, and 114 who chose not to disclose their gender.

\subsection{Factors measuring legitimacy}

There is a need for policymakers to empirically assess the legitimacy of their decision making within different contexts. This is particularly true for the practice of participatory budgeting, as achieving legitimacy is one of the main goals of this type of process. Therefore, in this section, we present legitimacy as a factor to distinguish between neighbourhoods and projects that have decisive and indecisive preferences, which can aid in optimising the participatory budgeting process.

%\begin{figure}
%  \includegraphics[width=0.5\textwidth,height=7.5cm]{Figures/proposed_projs.eps}
%   \caption{Number of proposed projects related to different sectors}
%  \label{fig:proposed_projs}
%\end{figure}

Inspired by a metric of dispersion in power systems (load factor) that measures power peak load, which can yield blackouts~\cite{Nambi2016,Mashlakov2021}, we introduce an extended and inverse version of this metric in Eq.~\ref{eq:legit} to calculate the legitimacy of selecting the \textit{top-k} project sector(s) or neighborhoods to fund:

\begin{equation}
   \begin{split}
        L=\frac{\sum_{i=1}^K Top_i}{\mu}. %\times \frac{1}{ \sum_{i=1}^N \frac{P_i}{\mu}}
   \end{split}
   \label{eq:legit}
\end{equation}

%\begin{equation}
%   \begin{split}
%        L=\frac{\frac{\sum_{i=1}^K Top_i}{\mu}}{\sum_{i=1}^N \frac{P_i}{\mu}}
%   \end{split}
%   \label{eq:legit}
%\end{equation}

\noindent In the equation, $\sum_{i=1}^K Top_i$ represents the total number of preferred project proposals or neighborhoods related to ${Top_k}$ chosen project sector(s) or neighborhoods to do investment in. The project sectors are derived from Question 26 in Table \ref{tab:q_cat}. The $\mu$ is the mean value of the number of the project sectors or neighborhoods respectively.  %$N$ represents the total number of project sectors,  and $P_i$ represents the number of proposed projects related to the project sector $i$.

Intuitively, the legitimacy metric calculates how well the top-k project sectors or neighborhoods represent a strong majority, i.e. how steep the peak of the top-k projects sectors are. For instance, when the population is indecisive, the top-k project sectors do not distinguish well from the rest of the project sectors; as a result, investing in them comes with a legitimacy risk by leaving a large portion of the population unsatisfied. 

Note that Equation~\ref{eq:legit} serves two legitimacy use cases: (i) investment prioritization to \emph{specific projects} for a certain neighborhood and (ii) investment prioritization to \emph{specific neighborhoods} for a certain project. For instance, in the context of participatory budgeting, the first use case supports the legitimacy of the voting outcomes, while the second one can resolve proportionality challenges~\cite{Peters2021} by determining legitimate neighborhoods where a project should be implemented or voting campaigns can be performed. 

%Our legitimacy calculation can be used within different contexts. As such, we also present legitimacy as a factor to highlight the optimal number investments within a particular project type across sectors, which may aid policymakers' top-down investment strategy and the design of participatory interventions. In doing so, $\mu$ is the mean of total number of responses for this project divided by the total number of neighborhoods. $\sum_{i=1}^K Top_i$ represents the sum of responses for the top $i$ neighborhood, with $i$ varying from 1 to 10 in our case. %${Top_k}$ represents the chosen neighborhood(s) where the project is to be funded, and $N$ represents the total number of neighborhoods. 

\subsection{Factors for quality of life and participation to predict satisfaction}

	\begin{table*}[!htb]
\caption{Survey questionnaire, dependent and independent variables}
\centering
\begin{footnotesize}

\begin{tabular}{p{0.5cm}p{8.5cm}}
\toprule
\multicolumn{1}{c}{\textbf{QID}} & \multicolumn{1}{c}{\textbf{QUESTIONS}}                                      \\ \midrule
\multicolumn{2}{c}{\bf Dependent Variable}\\
\multicolumn{2}{p{11cm}}{{\bf Encoding}: 4 (Very Good)|| 3 (Good)|| 2 (Enough)|| 1 (Insufficient)}\\
1                                & Quality of Life in Aarau                                             \\ \hline

\multicolumn{2}{c}{\bf Independent Variables}\\
\multicolumn{2}{c}{\bf Satisfaction Sectors $\rightarrow$ Quality of Life Indicators}\\
\multicolumn{2}{p{12cm}}{{\bf Encoding}: 5 (Very Satisfied)|| 4 (Satisfied)|| 3 (Neutral)|| 2 (Not Satisfied)|| 1 (Not at All Satisfied) || 0 (I don’t know) }\\
2                                & Social meetings facilities satisfaction                                     \\
3                                & Neighbourly life satisfaction                                               \\
4                                & Shopping facilities satisfaction                                            \\
5                                & Social infrastructure satisfaction                                          \\
6                               & Housing environment satisfaction                                            \\
7                               & Footpath network satisfaction                                               \\
8                               & Bike path network satisfaction                                              \\
9                               & Public transport satisfaction                                               \\
10                               & Recreational areas satisfaction                                             \\
11                              & Playing facilities satisfaction                                             \\

12                               & Security satisfaction                                                       \\ \hline
\multicolumn{2}{c}{\bf Independent Variables}\\

\multicolumn{2}{c}{\bf Participation of citizens in social activities $\rightarrow$ Participation Indicators}\\
\multicolumn{2}{p{11cm}}{{\bf Encoding}:  4 (Daily)|| 3 (Weekly)|| 2 (Monthly)|| 1 (Less often) || 0 (I don’t know my neighbours) }\\

13                                & How often did you get involved in your neighbourhood in the last month?                                                   \\
14                                & How often do you get in contact with your neighbours?                                        \\
\multicolumn{2}{p{11cm}}{{\bf Encoding}:  2 (Yes)|| 1 (Maybe) || 0 (No) }\\

15                               & Would you like to get more involved in your neighbourhood?                                        \\

16                              & Do you wish there was more neighbourly contact?                                      \\
17                              & Would you like increased sharing offers in your neighbourhood?                                       \\ 
\multicolumn{2}{p{11cm}}{{\bf Encoding}:  2 (High)|| 1 (Medium) || 0 (Low) }\\
18                              & How many people do you approximately know in your neighbourhood? \\ \hline
\multicolumn{2}{c}{\bf Miscellaneous, Demographic}\\
\multicolumn{2}{p{11cm}}{Free Text }\\

19                               & In which neighbourhood do you live?                                         \\
20                               & Did you live in another neighbourhood before? \\
21                                & In which neighbourhood did you live?                                        \\ 
\multicolumn{2}{p{11cm}}{{\bf Encoding}:  1 (Alone)|| 2 (Household with Children)|| 3 (Household for two)|| 4 (Shared) }\\
23 & How do you live? (Household type)\\
\multicolumn{2}{p{11cm}}{{\bf Encoding}:  1 (Basic Vocational)|| 2 (Compulsory School)|| 3 (higher technical college)|| 4 (University degree) }\\
24 & What is your highest educational attainment?\\
\multicolumn{2}{p{11cm}}{{\bf Encoding}:  0 (Not Working)|| 1 (Retired) || 2 (Part Time)|| 3 (Full Time) }\\
25 & What is your current employment status?\\

26 & Project ideas \\ \hline

\end{tabular}
				\label{tab:q_cat}
\end{footnotesize}
\end{table*}

\begin{figure*}[!htb]
\centering

 \includegraphics[scale=0.5]{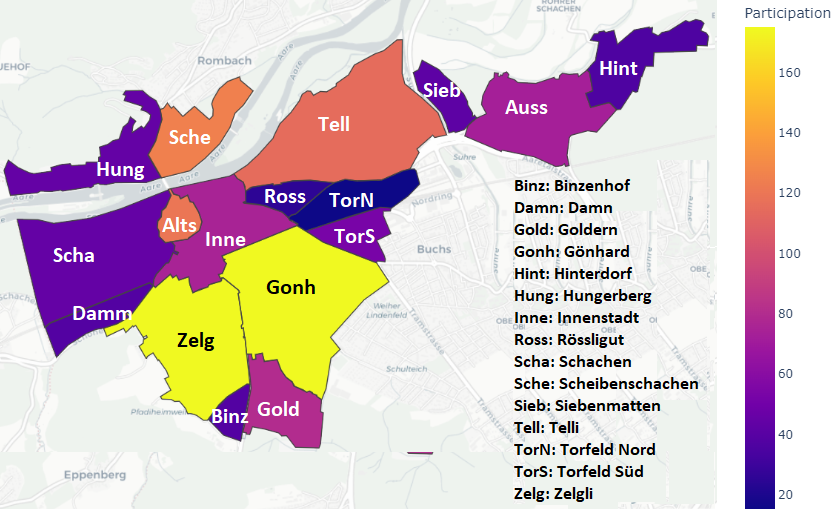} 
  \caption{Number of survey participants in different neighborhoods of Aarau.}
  \label{fig:p1}
\end{figure*}

 The survey in Aarau was designed with questions to extract the satisfaction of people regarding the various project sectors (Question 2-12, Table~\ref{tab:q_cat}), their social connections and participation within their neighbourhood (Question 13-18, Table~\ref{tab:q_cat}) along with demographic information such as place of stay, the form of commute, housing conditions, etc. (Question 19 -26, Table~\ref{tab:q_cat}). Finally, the participants also provide an overall quality of life score.

 Fig~\ref{fig:resp_from_each_neighborhood} shows that the participation in the survey varied across neighbourhoods. For instance, \textit{Zelgi} has the highest number of respondents (175 participants) followed by \textit{Gönhard} (173 participants). In contrast, the neighbourhood \textit{Torfeld Nord} has the lowest number of participants (7 participants) followed by \textit{Rössligut} (20 participants) (refer to the map in Fig.~\ref{fig:p1}). Overall, 520 participants marked quality of life in Aarau as `Very Good' and 592 participants marked quality of life as `Good'. Only 8 participants marked the quality of life in Aarau as `Bad' whereas 6 participants marked `I don\'t know' as an answer to the question about the quality of life in Aarau.

\subsection{Classification model}\label{ssec:reg_model}

{\em Dependent and independent variables}: We introduce a classification model to explain the people's perception of the overall quality of life (dependent variable) using the responses in the questions related to their satisfaction in individual project sectors such as shopping facilities, sports infrastructure, recreational spaces, bike path, transport, etc (independent variables). As demonstrated in literature, satisfaction with public facilities is related to perceptions of quality of life \cite{leonard2009, sirgy2009community}. Furthermore, we also use independent variables related to participation such as the people's participation in community events, neighbourly contacts, etc. Table~\ref{tab:q_cat} enumerates the various satisfaction and participation factors we use in the classification model to interpret the overall citizens' quality of life. 

The  responses were ordered categorical variables and we encoded them for numerical representation (details of encoding in Table~\ref{tab:q_cat}). The data was highly imbalanced among the various classes of quality of life and hence we combine responses related to `bad', `I don\'t know' and `Insufficient' as a single class representing perceived quality of life as `Insufficient'. 

While dealing with ordered categorical variables, an ordinary least squared regression model may yield inaccurate and misleading results. Employing a classification model in a supervised setup (as ground truth is known), we identify the significant independent variables, which may play an important role to predict the overall citizens' quality of life in Aarau. We are dealing with a total of 17 factors (satisfaction and participation factors), 4 classes and 1,204 data points with highly uneven data distribution among classes. Hence, empirically, we experimented with the logit regression model~\cite{hosmer2013applied} and decision tree model~\cite{song2015decision}, however, the results are overfitted and inconclusive. This is because the non-linear relationships between dependent and independent variables need to be projected into a higher dimensional space to extract a better decision boundary so that the dependent variable falls into each of the classes.

We use polynomial approximations for better prediction of probability in multiple classes, and employ support vector machines~\cite{meyer2015support} and a 2-layer neural network~\cite{gurney2018introduction} for determining the significant independent variables. Both of the models provide higher accuracy and as the data is highly imbalanced among classes, the neural network is superior.

The output probability of a class (i.e. $class_i$) is calculated as:

\begin{equation}
    output(class_i) = g_o({\sum\limits_{{n}\in N}}{{g_n  ({\sum\limits_{{j}\in J}}{I_j\cdot W_j} + 1})\cdot W_n} + 1),
    \label{eq:class}
\end{equation}

\noindent where $J$ denotes the input dimension (the factors), $N$ denotes total number of learning neurons, $g_o$,  $g_n$ are functions that decide which learning neurons get activated in every layer to analyse and approximate the input data to deduce the probability of being below or at class ${i}$. The approximation process takes place in two layers. $W_n$ denotes the initial weight (preference) matrices for neurons in every layer. 

\subsection{Citizens' relocations for improving quality of life}

In this section, we propose a metric to calculate the quality improvement or deterioration (in terms of the satisfaction level of facilities and the services provided in the neighbourhoods) while moving from one neighbourhood to another. Considering a scenario of relocation from a neighbourhood ${x}$ to neighbourhood ${y}$, the \emph{Relative Quality Improvement} (${RQI_{x,y}}$) is  calculated using Equation \ref{eq:rqi}:

\begin{equation}
    RQI_{x,y}= \frac{1}{k}\sum_{j=1}^{k}{\frac{\overline{Q}_{j,y}-\overline{Q}_{j,x}}{\overline{Q}_{j,x}}}.
    \label{eq:rqi}
\end{equation}

\noindent In the above equation, ${\overline{Q}_{j,y}}$ represents the mean satisfaction of the project type ${j}$ in the neighbourhood ${y}$, ${\overline{Q}_{j,x}}$ represents the mean satisfaction of the project type ${j}$ in the neighbourhood ${x}$, and ${k}$ represents the total number of projects under consideration. Positive $RQI$ represents improvement by the relocation, while the negative $RQI$ represents a deterioration. 

Furthermore, we also introduce a metric to measure for each individual and project sector the \emph{Perceived Quality Improvement} (PQI) by a relocation from the neighbourhood ${x}$ to neighbourhood ${y}$:

\begin{equation}
    PQI_{x,y}=1-\frac{\overline{Q}_{j,y}-Q_{j,y}}{\overline{Q}_{j,y}-\overline{Q}_{j,x}}.
    \label{eq:pqt}
\end{equation}

\noindent In the above equation, $\overline{Q}_{j,y}$ represents people's mean satisfaction level about project sector $j$ in neighbourhood $y$, $Q_{j,y}$ represents the individual's satisfaction level about the project type $j$ in neighbourhood $y$, and $\overline{Q_{j,x}}$ represents people's mean satisfaction level about the project type $j$ in the neighbourhood $x$ (neighbourhood from where the person relocated).

RQI measures whether a relocation of an individual improves quality of life based on the satisfaction level of all residents in neighbourhoods, while PQI measures in a more personalized way the perceived improvement in quality of life (relative to the collective one).

\section{Evaluation and Results}\label{sec:Results}

In this section, we provide the results of the experiments we conducted to answer $RQ1, RQ2$, and $RQ3$.

\subsection{RQ1: How to prioritize bottom-up participatory interventions to improve legitimacy of policy-making?}

\begin{figure*}[!htb]
\centering
\begin{minipage}{.5\textwidth}
 \includegraphics[scale=0.15]{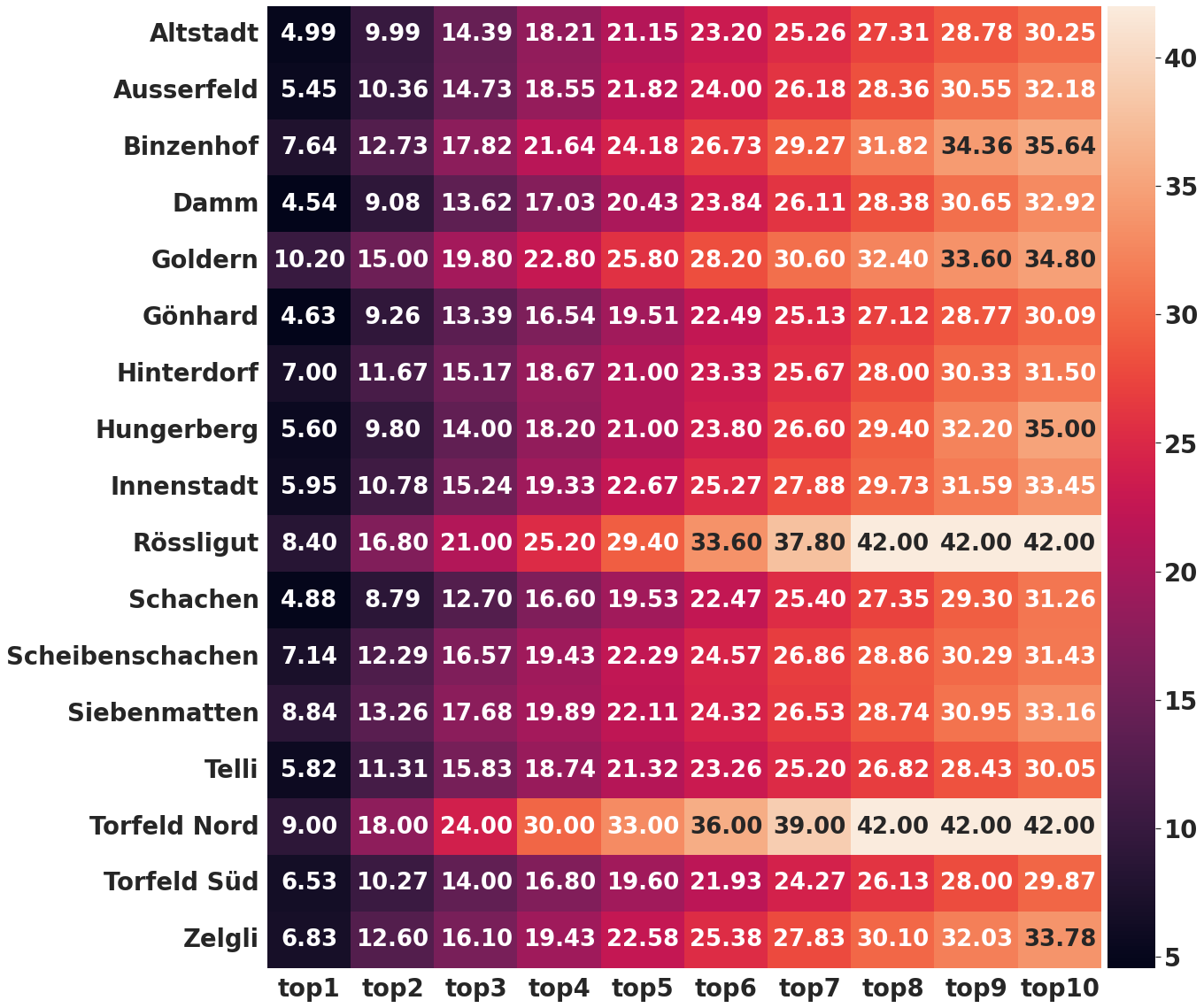} 
  \captionof{figure}{Legitimacy [\%] for top-k project sectors per neighbourhood (using Eq.~\ref{eq:legit}).}
  \label{fig:leg_lf}
\end{minipage}%
\begin{minipage}{.5\textwidth}
  \includegraphics[scale=0.5]{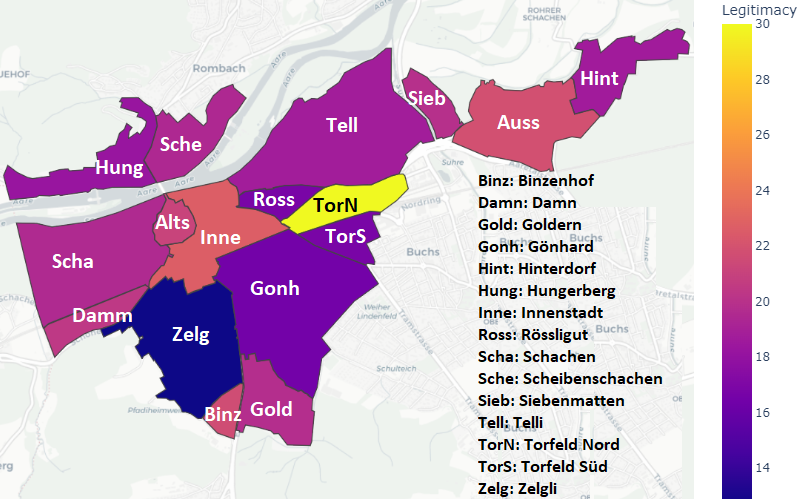}
  \caption{Legitimacy map for optimal top-k of project sectors in different neighbourhoods of Aarau.}
  \label{fig:l1}
  \end{minipage}
\end{figure*}

\begin{figure}[!htb]
  \includegraphics[scale=0.47]{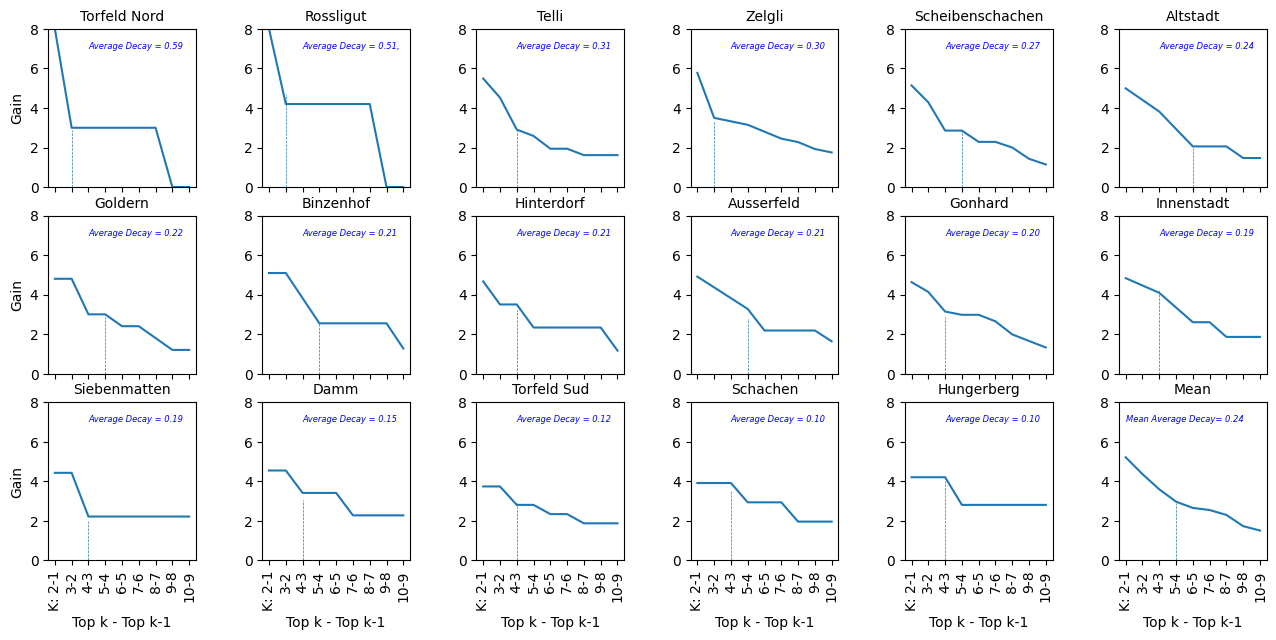}
   \caption{Decay of legitimacy gain as top-k project sectors increase. The vertical lines depict the optimal $k$ projects (knee point) sorted by average decay rate.}
  \label{fig:leg_gain}
\end{figure}

Our legitimacy calculation offers a solution for policymakers when seeking to optimize decision-making processes. Specifically, it offers guidance regarding which neighborhoods might benefit from participatory interventions to bolster legitimacy. We present our results in Figure~\ref{fig:leg_lf} to illustrate the legitimacy levels attainable if a specific number of project sectors are funded per neighborhood ($k$). To calculate the optimal-k scores, we compare the potential legitimacy gains as the number of project sector investments increases, identifying the point where the increase in legitimacy begins to decay (using the elbow method to make this determination~\cite{cui2020introduction}). The optimal-k, represented in Figure~\ref{fig:leg_gain} as the dotted vertical line within each plot of the neighborhood, can be thought of as the \emph{minimum investment for the highest returns}. Within this framework, a lower optimal-k score signifies that a local authority can maximize the legitimacy levels with minimal investment. In contrast, a high optimal-k suggests that a local authority shall finance a greater number of project sectors to achieve the highest return in legitimacy (which may be challenging when the authority does not have the funding capacity for such investments). As such, a neighborhood with a higher optimal-k may be deemed suitable for a participatory intervention to compensate for the inability of the local authority to maximize legitimacy returns.

Fig.~\ref{fig:leg_gain} demonstrates how \textit{Altstadt} has the highest optimal k-value, suggesting that the minimum investment for the highest return is present when six project sectors are funded. In addition, \textit{Scheibenschachen, Goldern, Binzenhof} and \textit{Ausserfeld} have all been determined as optimal at five project sectors. As such, the aforementioned neighborhoods may be considered as the most likely candidates for participatory intervention due to the increased cost associated with the minimum level of investment. \textit{Torfeld Nord} and \textit{Zelgli} have the lowest optimal-k value, indicating that a minimum investment in two project sectors yields the highest return. As such, these areas may gain less from participatory interventions, as local authorities can more readily meet funding needs. 

% \begin{figure}[!htb]
%   \includegraphics[width=0.5\textwidth]{Figures/map.png}
% \caption{How Top-k projects popularity impact the neighbourhoods}
% \label{fig:topk}
% \end{figure}

In Fig.~\ref{fig:leg_gain}, we also identify how the legitimacy gain decays as the number of project sectors selected increases. For example, the data suggest that in \textit{Rössligut}, the legitimacy gain plateaus after funding two sectors, with zero increase beyond eight project sectors. In addition to offering guidance regarding participatory interventions, these findings can inform funding allocation decisions by policy makers, who may choose to redirect resources from areas such as \textit{Rössligut} to neighbourhoods that see more of a significant increase in legitimacy should more project sectors be funded, such as \textit{Damm} or \textit{Hungerberg}. As such, this approach optimizes the allocation of resources to maximize legitimacy outcomes across neighbourhoods.

\begin{figure}[htbp]
  \centering
  \includegraphics[width=0.6\textwidth]{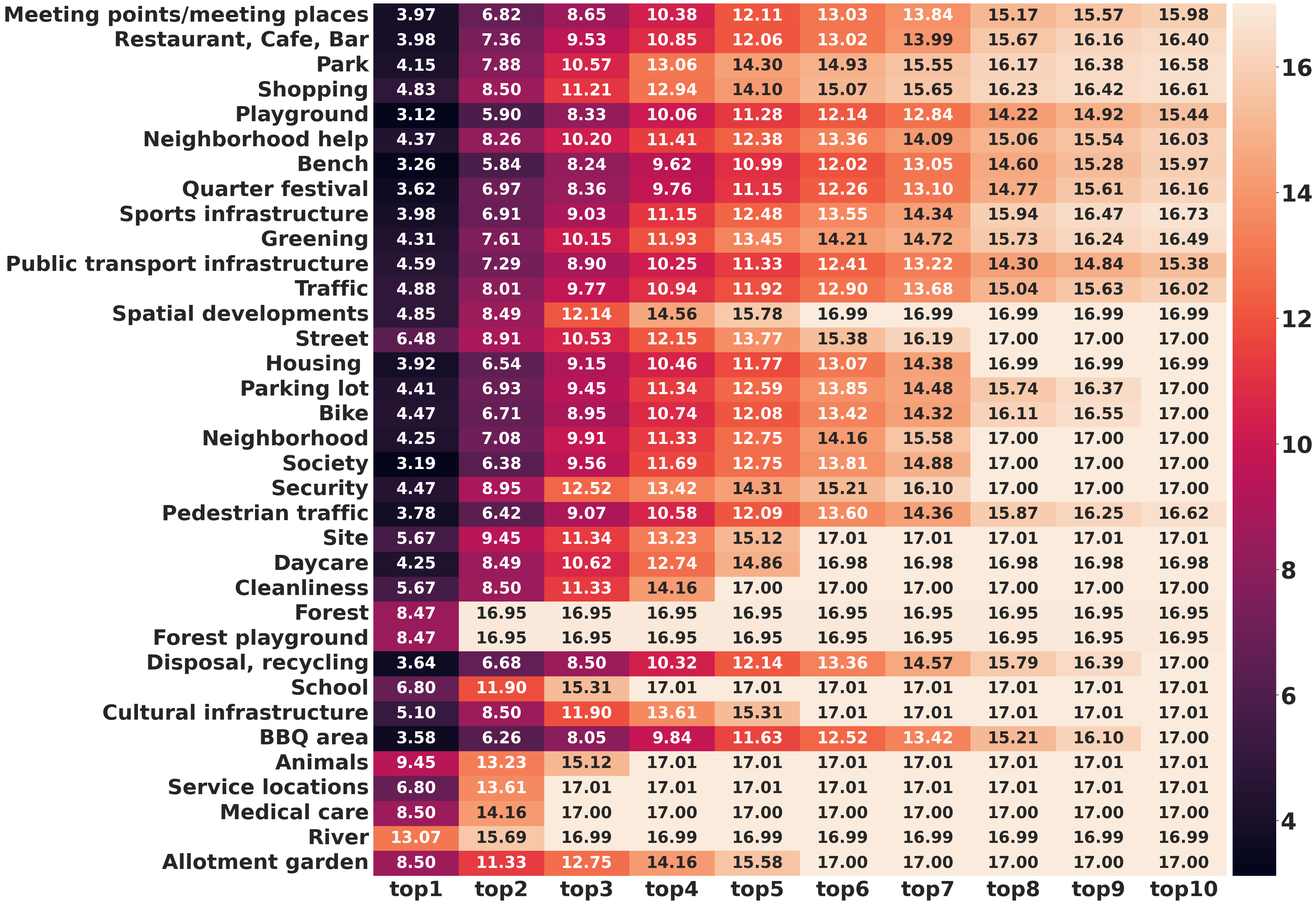}
  \caption{Legitimacy [\%] for top-k neighbourhoods per project type (using Eq. 1).}
  \label{fig:projl}
\end{figure}

\begin{figure}[htbp]
  \centering
  \includegraphics[width=0.97\textwidth]{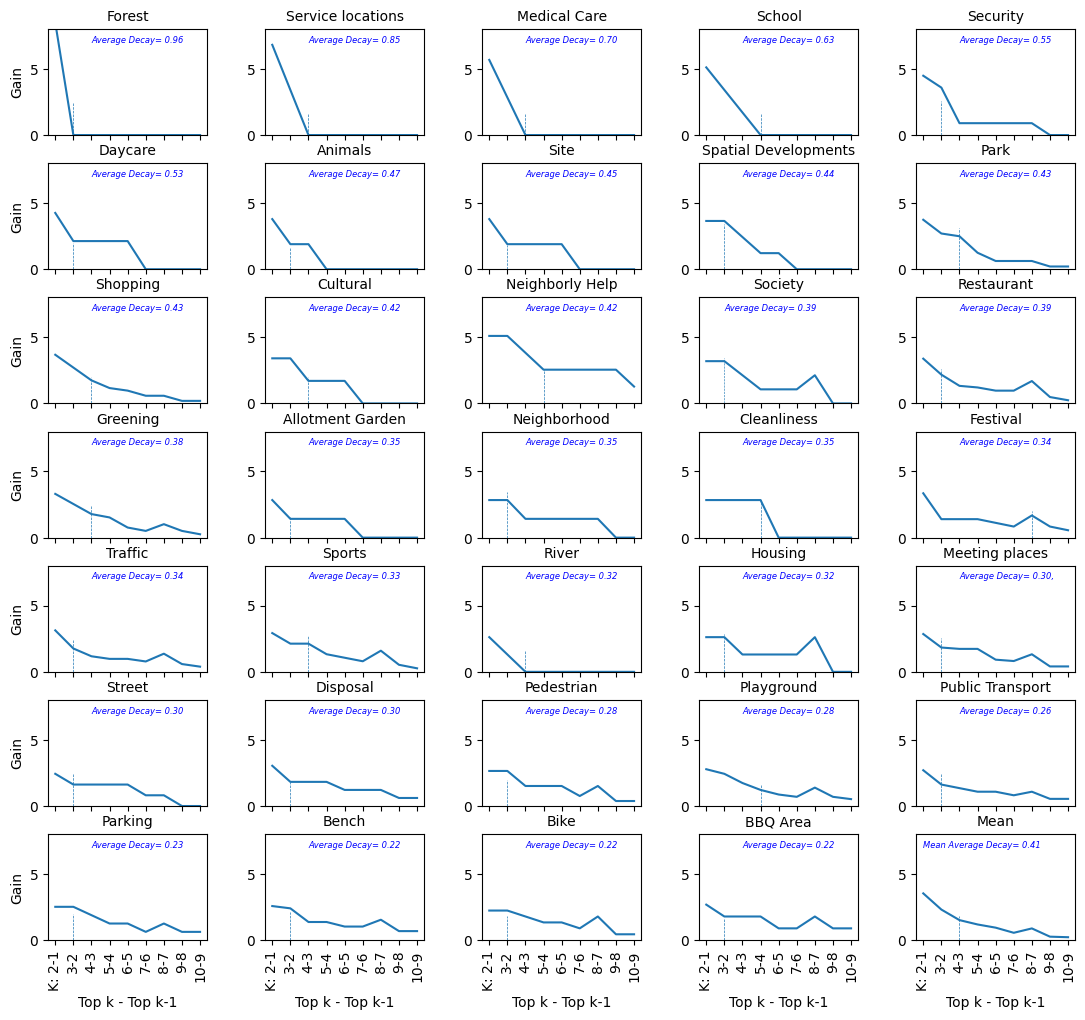}
  \caption{Decay of legitimacy gain as top-k neighbourhoods increase. The vertical lines depict the optimal $k$ projects (knee point). Sorted by average decay rate.}
  \label{fig:proj_g}
\end{figure}

Figure~\ref{fig:projl} illustrates the legitimacy gain that policymakers could achieve by funding projects across various neighborhoods, providing insights into the optimal number of investments. For instance, investing in `Rivers' within a single neighborhood yields a significant return in legitimacy, with a smaller increase should the authority invest in `Rivers' in two or more neighborhoods. However, if the local authority aims to fund `Playgrounds', our metric suggests that it should do so across a larger number of neighborhoods to maximize legitimacy returns. 

\begin{figure*}[!htb]
\centering
\begin{minipage}{.5\textwidth}
 \includegraphics[width=1.0\textwidth]{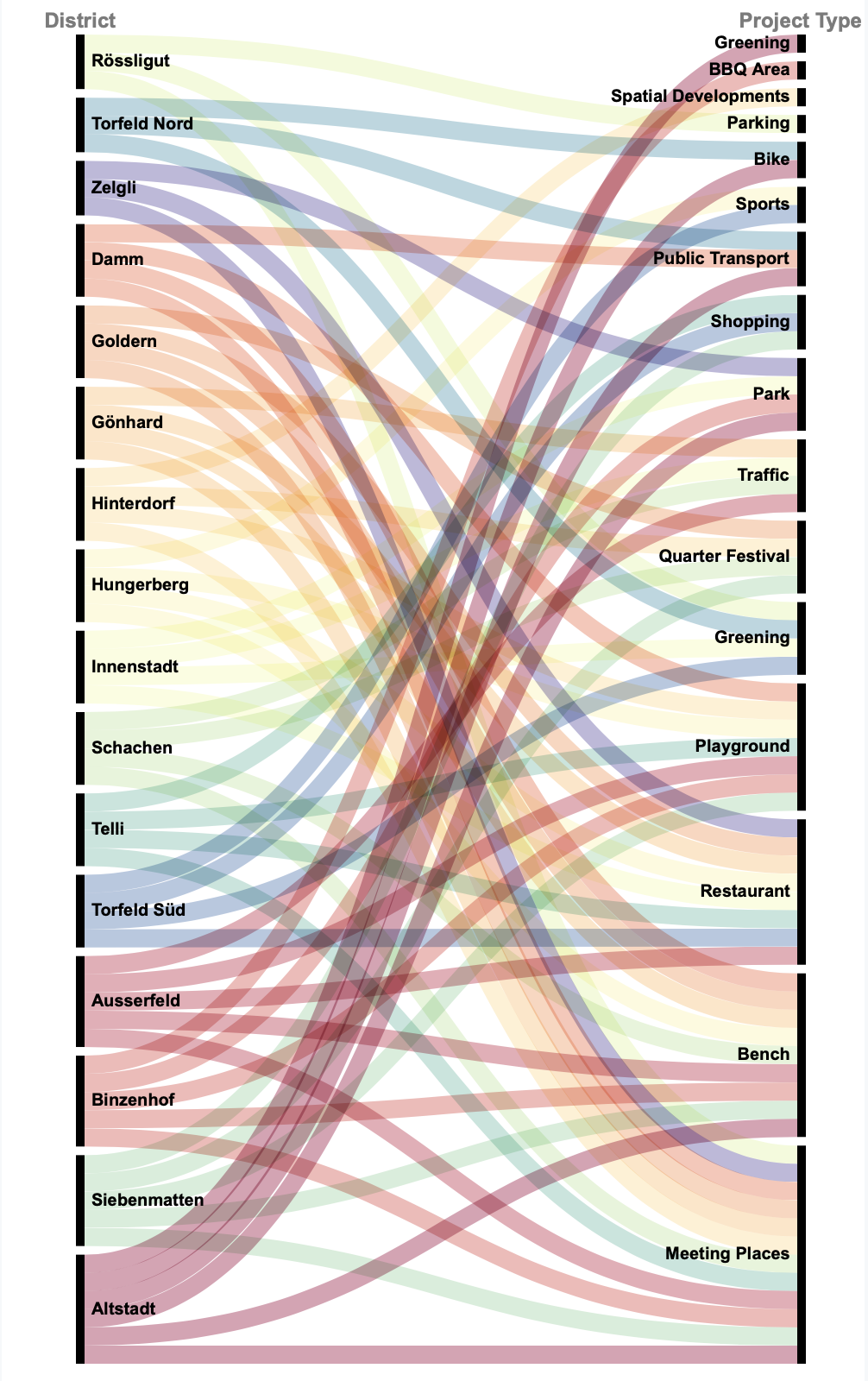} 
  \caption{Projects with legitimacy gain for each different neighbourhood.}
  \label{fig:distsan}
\end{minipage}%
\begin{minipage}{.5\textwidth}
  \includegraphics[width=1.0\textwidth]{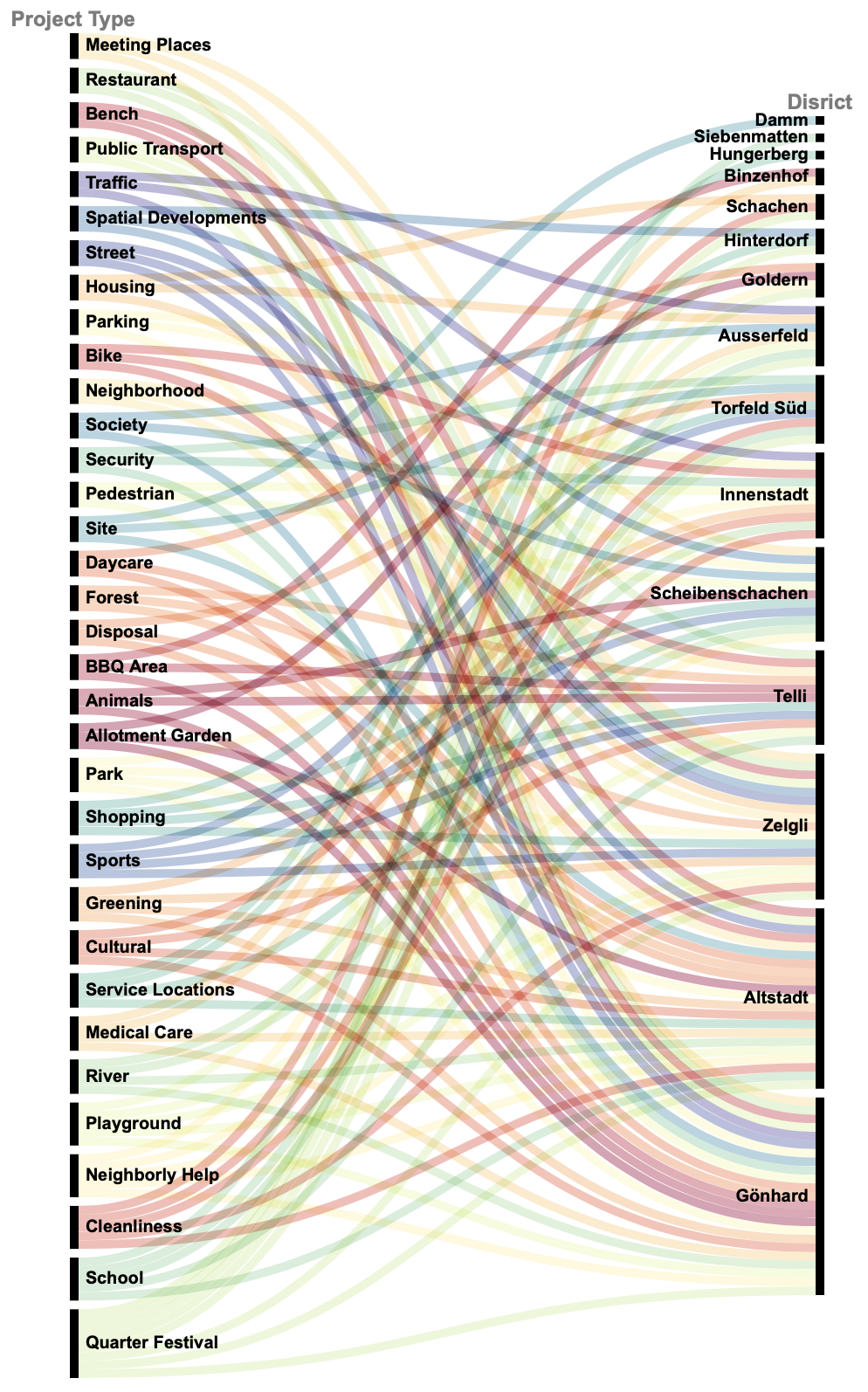}
  \caption{Neighbourhoods with legitimacy gain for each different project.}
  \label{fig:prosan}
  \end{minipage}
\end{figure*}

Fig.~\ref{fig:distsan} illustrates the specific optimal-k project types per neighborhood with legitimacy gain. It is observed that the optimal-k project sectors in \textit{Rössligut} is three. Within \textit{Rössligut}, Fig.~\ref{fig:distsan} demonstrates how the greatest legitimacy gain comes from investments in `Parking', `Greening' and `Meeting Places'. Fig.~\ref{fig:distsan} can be beneficial for the design of participatory budgeting, particularly filtering project proposals before the voting procedure. Filtering is a challenge for policymakers (particularly when conducted without bottom-up consultation), as it risks being viewed as a discretionary practice aimed at selecting projects that conform with the authorities preexisting policy direction \cite{aleksandrov2018participatory, allegretti2014lisbon}. 

In Fig.~\ref{fig:proj_g}, we move further by identifying the optimal-k investments per project type and the legitimacy decay as the number of neighbourhoods increases. For example, `Neighbourly Help' is optimal to invest on within five neighbourhoods. Further investment in `Neighbourly Help' surpasses the optimal legitimacy point, Fig.~\ref{fig:proj_g} demonstrates that the legitimacy return is relatively negligible. Fig.~\ref{fig:prosan} presents the specific optimal-k neighborhoods per project type for legitimacy. For example, we previously determined that the optimal-k investment for `Parks' stood at four neighborhoods. Specifically, in Fig.~\ref{fig:prosan}, we can see that the greatest legitimacy gain comes for 'Park' investments in the locations of \textit{Zelgli, Gönhard, Altstadt} and \textit{Innenstad}. We can also distinguish projects, whose legitimacy gain is made by several different neighborhoods such as `Cleanness' and `School' and `Neighborhood Festival'.

%In this sense, this metric is beneficial for policymakers as it can offer guidance in terms of the degree of investment per project type, empowering policymakers to make targeted investments and providing empirical guidance for allocating funds. 

\subsection{RQ2: Which indicators of satisfaction and participation explain citizens’ overall quality of life?}

We design a classification model that offers policymakers further guidance regarding the type of investment that benefits citizens quality of life. As the data is highly imbalanced within classes, we employ sampling to create synthetic data for minority classes.  We use the method proposed in~\cite{chawla2002smote} to join minority class data points with their nearest neighbours and use various distance functions to generate additional data points. We demonstrate the accuracy obtained for predicting the significant independent variables in Table~\ref{tab:static1}. We achieve an overall accuracy of 74.3\% with synthetic sampling data, considering all satisfaction and participation factors. This is significantly higher than the dataset without sampling. The satisfaction factors have a higher impact and are more significant in predicting quality of life than participation factors.

\begin{table}[!htb]
\centering
\caption{Relationship between the most popular project sectors and their statistical significance in the classification model that explain citizens' overall satisfaction. Note that 5/7 project sectors come with $p<0.05$, which validates the designed legitimacy measure. We also include here for further validation the projects with the optimal legitimacy gain.}
\label{tab:secqol}
\begin{tabular}{p{4cm}p{1.5cm}p{4.0cm}p{4.3cm}}
\toprule
Project sectors	&	{\bf Ranking}	&	{\bf Optimal Legitimacy Gain [\%]} & {\bf Classification model p-values}	\\ \hline
Social Facilities / Meetings	&	1 & 18.13	&	0.019	\\ 
Recreational Spaces	&	2	& 13.8	 & 0.001	\\ 
Playing Facilities	&	3	&	10.8 & 0.001	\\ 
Social Infrastructure	&	4	&	13.21 & 0.018	\\ 
Public Transport	&	5	&	 .05 & \textcolor{red}{0.358}	\\ 
Shopping Facilities	&	6	& .03	& \textcolor{red}{0.528}	\\ 
Neighbourly Life	&	7	& .07 &	0.001	\\ 
\bottomrule
\end{tabular}
\end{table}

 All independent variables, \textit{neighbourly life, housing environment, footpath network, recreational areas} and \textit{security}, apart from \textit{public transport} and \textit{shopping facilities} appear to be significant in terms of their relationship with citizens' overall quality of life assessment and have p-values lower than 0.05. Using only participation factors does not better predict the quality of life and provides lower accuracy. Combining both factors gives the highest overall performance even though the p values of the participation factors are mostly not significant (i.e $p>0.05$). In Table~\ref{tab:secqol} the `Ranking' column highlights the number of times citizens proposed projects relating to such project sectors (e.g., the highest number of proposals is related to social facilities/meetings). In the 'Optimal Legitimacy Gain [\%]' column, the percentage of such projects is found within the mean optimal k for all neighborhoods, which also validates the 'Ranking' column. This measurement is beneficial as it may offer guidance for policymakers in terms of the project types that may return the greatest level of legitimacy.

\begin{figure*}[!htb]
\centering
\begin{minipage}{.45\textwidth}
 \includegraphics[scale=0.34]{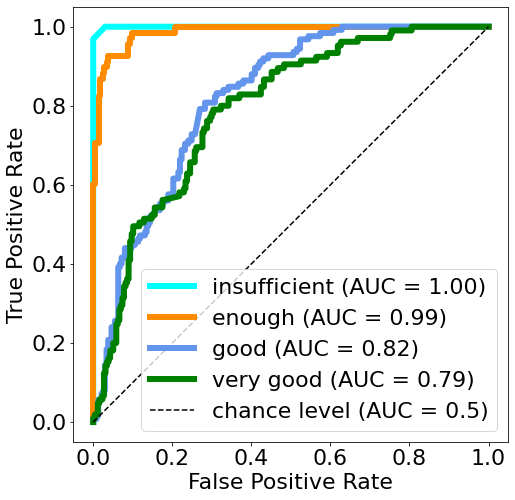} 
  \captionof{figure}{Distribution of the true and false positive prediction of quality of life - all classes, (AUC: Area Under the Curve represents the probability of correct prediction).}
  \label{fig:tpr1}
\end{minipage}
\begin{minipage}{.45\textwidth}
  \includegraphics[scale=0.34]{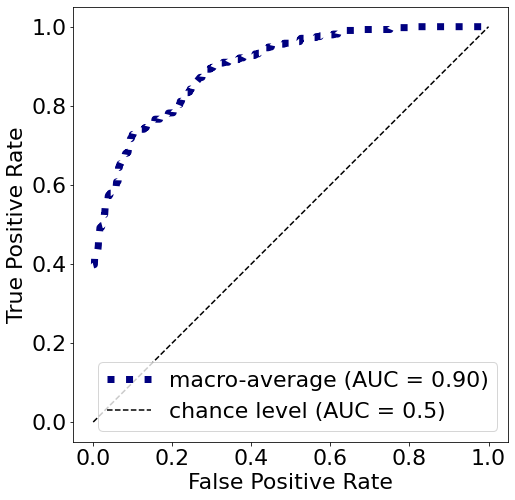}
  \captionof{figure}{Distribution of the true and false positive prediction of quality of life - average of classes.}
  \label{fig:tpr2}
  \end{minipage}
\end{figure*}

 \begin{table*}[!ht]
\caption{Characterization of the results}
\centering
\begin{scriptsize}
\begin{tabular}{p{1.1cm}p{4.14cm}rrrrrrrrr}
\hline
\multicolumn{1}{p{1.17cm}}{\bf Dataset} & \multicolumn{1}{c}{\bf Quality of Life} & \multicolumn{3}{c}{\bf Satisfaction (S)} & \multicolumn{3}{c}{\bf Participation (P)}  & \multicolumn{3}{c}{\bf S + P} \\ \hline

\multicolumn{11}{p{12cm}}{cross entropy loss, adam optimiser, drop out = 0.5 in last  layer, 2 layer dense neural net, Leaky Relu}\\ \hline

\multicolumn{2}{c}{} & \multicolumn{1}{p{0.4cm}}{\bf {\tt \bf r}} & \multicolumn{1}{p{0.4cm}}{\bf {\tt \bf  p}}  & \multicolumn{1}{p{0.4cm}}{\bf {\tt \bf Acc}} &  \multicolumn{1}{p{0.4cm}}{\bf {\tt \bf r}} & \multicolumn{1}{p{0.4cm}}{\bf {\tt \bf p}}  &  \multicolumn{1}{p{0.4cm}}{\bf {\tt \bf Acc}} & \multicolumn{1}{p{0.47cm}}{\bf {\tt \bf r}} & \multicolumn{1}{p{0.4cm}}{\bf {\tt \bf p}}  & \multicolumn{1}{p{0.4cm}}{\bf {\tt \bf Acc}} \\ \hline

\multirow{4}{*}{\parbox{1cm}{Without Sampling }} &  1 (Insufficient) &      0.001 &     0.002   &   0.001  &        0.125  &   0.200  &   0.124    &  1.000   &  0.200  &   \textcolor{green}{1.000}  \\ 
 &          2 (Enough)  &    0.167  &    0.091  &    0.166    &     0.010  &   0.015   &  0.016      & 1.000   &  0.182    &  \textcolor{green}{0.308} \\
 &          3 (Good)   &   0.520   &   0.731   &   0.512    &   0.507   &  0.694   &  0.506     &  0.576  &  0.769  &   \textcolor{green}{0.659} \\
  &         4 (Very Good)   &    0.554   &   0.330 &     0.553     & 0.452  &   0.298  &   0.411      & 0.606  &   0.457   &  {\bf \textcolor{green}{0.605}} \\
  
& {Overall metrics} & {\bf 0.310} & {\bf 0.288} & {\bf 0.509} & {\bf 0.271} & {\bf 0.298} & {\bf 0.477} &  {\bf 0.796}    &  {\bf 0.402} & {\bf 0.592} \\ \hline

  \multirow{4}{*}{\parbox{1cm}{Synthetic Data - Over Sampling }} &   1 (Insufficient) &       0.897   &   0.953  &    0.891  &        0.937    &  0.922    &  0.999         &   1.000     & 0.953    &  \textcolor{green}{0.999} \\
 &          2 (Enough)  &    0.875   &   0.824 &     0.872   &          0.782   &   0.632    &  0.781        & 0.894    &  0.868     & \textcolor{green}{0.893}   \\
 &          3 (Good)   &   0.674   &   0.496  &    \textcolor{green}{0.671}   &               0.529    &  0.584   &   0.523     &  0.626   &   0.776   &   0.624 \\
  &         4 (Very Good)   &  0.551  &    0.724   &   0.552     &             0.500   &   0.505    &  0.502      &  0.650   &   0.495    &  \textcolor{green}{0.656}  \\ 

& Overall Metric & {\bf 0.749} & {\bf 0.749} & {\bf 0.704} & {\bf 0.687} & {\bf 0.661} & {\bf 0.630}  & {\bf 0.792}  &   {\bf 0.773} & {\bf 0.748} \\ \hline
\multicolumn{11}{c}{r: Recall (\%); p: Precision(\%); Acc: Accuracy (\%)}\\
\end{tabular}

\label{tab:static1}
\end{scriptsize}

\end{table*}

In Figures~\ref{fig:tpr1} and~\ref{fig:tpr2}, we show the relation of the true and false positives ratios for the prediction model with which we obtain the highest accuracy. The \emph{Area Under the Curve} (AUC) metric helps to understand the probability of correct prediction of a data point to its true class and we see that the AUC is higher for the classes `Insufficient' and `Enough' (see classwise AUC in Figure~\ref{fig:tpr1}), however, the average AUC is 0.90 (Figure~\ref{fig:tpr2}) and hence we can conclude that the satisfaction factors can significantly help in estimating the overall citizens' quality of life.

\subsection{RQ3: Are citizens' relocation to different neighbourhoods associated with quality of life improvements?}

\begin{figure*}
\captionsetup{justification=raggedright}  
\centering
\begin{minipage}{.51\textwidth}
 \includegraphics[scale=0.2]{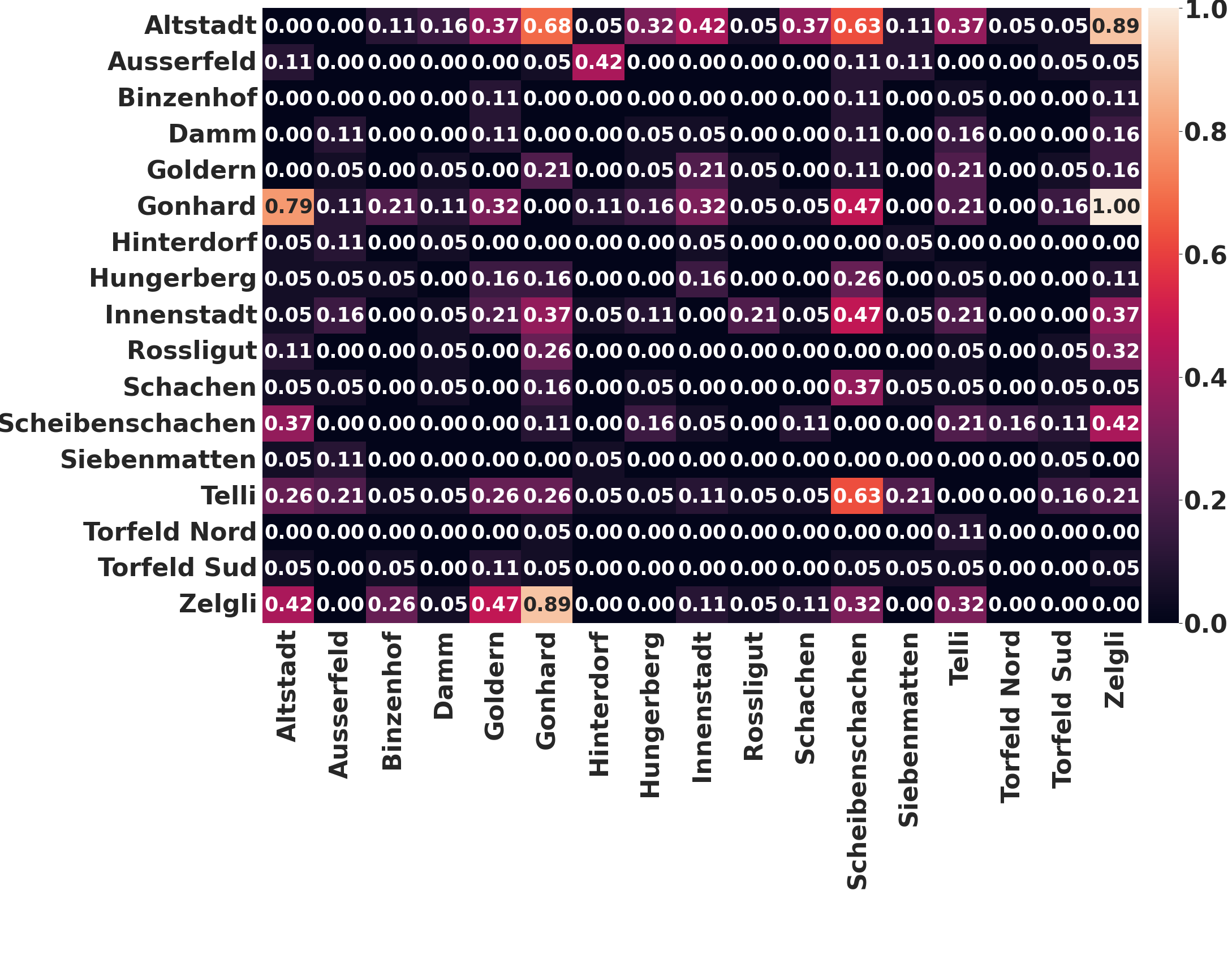} 
  \captionof{figure}{Normalized migration probability.}
  \label{fig:migration_stats1}
\end{minipage}
\begin{minipage}{.45\textwidth}
  \includegraphics[scale=0.201]{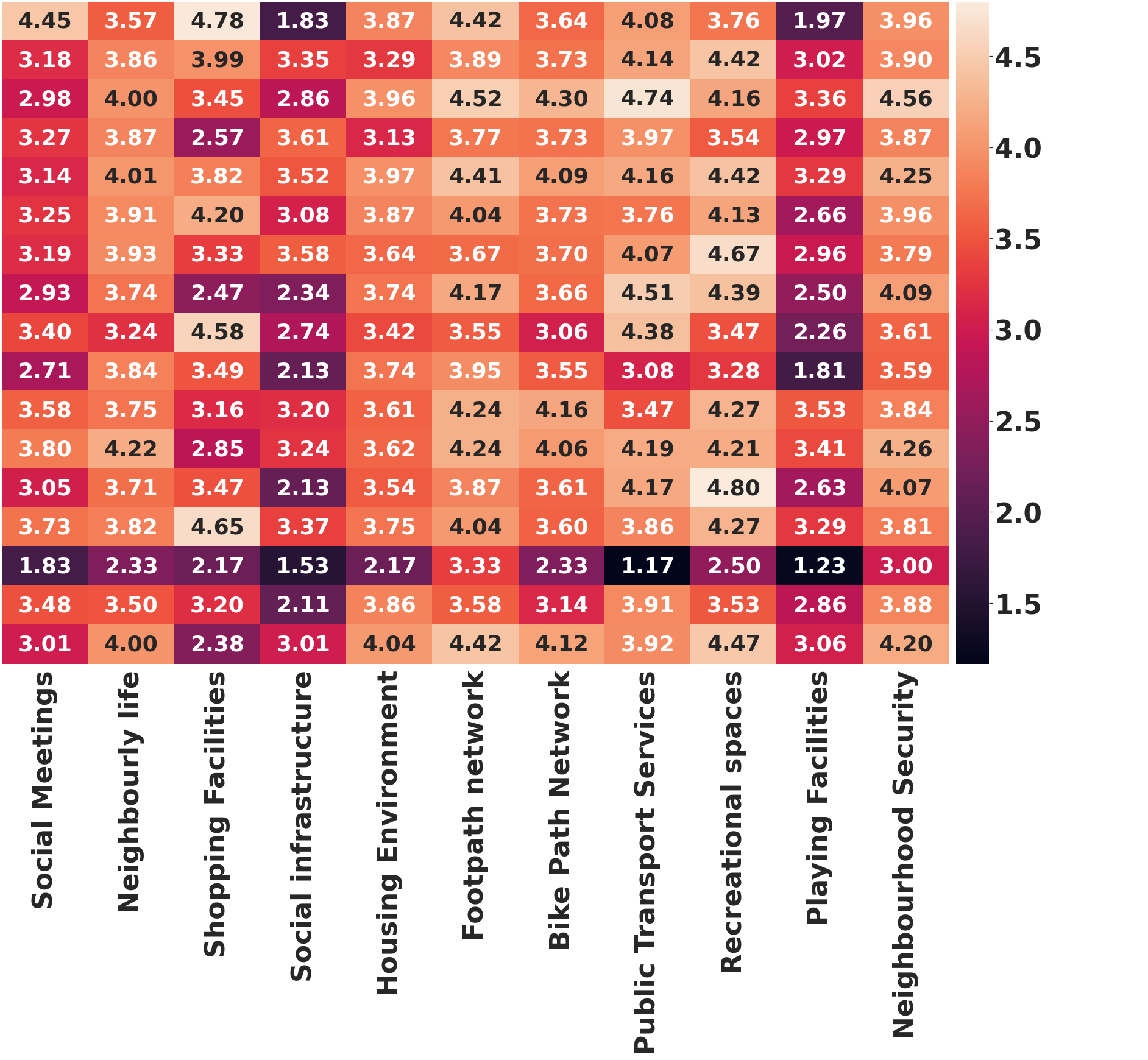}
  \captionof{figure}{Mean satisfaction.}
  \label{fig:msl_wrt_quest}
  \end{minipage}
\end{figure*}

To answer RQ3 we calculate the relocation probability (max-min normalized values) from one neighbourhood to another in Fig.~\ref{fig:migration_stats1}. There has been 19 migrations from Gonhard to Zelgli, which is the highest and also signifies the most likely migration people in Aarau can undertake. On the other hand migrations between some localities are very less, and often we find a singular instance.

Specifically, we investigate which project sectors citizens compromise on and which ones improve by relocating to a new neighbourhood, which can aid policymakers in understanding the value placed on project sectors during relocations. To do so, for each observed relocation, we calculate the overall $RQI$ and the $RQI$ for each project sector using the mean satisfaction level about the project sectors (Fig.~\ref{fig:msl_wrt_quest}). A positive $RQI$ value indicates a quality of life improvement based upon relocation, while a negative $RQI$ value indicates a deterioration.

 % \begin{figure*}
 % \centering
 % \begin{minipage}{.5\textwidth}

 % \includegraphics[scale=0.3]{Figures/migration.png}
 %   \caption{Migration probability within neighbourhoods}
 %  \label{fig:migration_stats}
 % \end{minipage}%
 % \begin{minipage}{.5\textwidth}
 % \includegraphics[scale=0.3]{Figures/satisfaction.png}
 %   \caption{Satisfaction Index of Neighbourhoods}
 %  \label{fig:msl_wrt_quest}
 %  \end{minipage}

 % \end{figure*}

% \begin{figure}[!h]
%   \includegraphics[scale=0.2]{Figures/migration.png}
%    \caption{Migration probability of moving from one neighbourhood to another}
%   \label{fig:migration_stats}
% \end{figure}

% \begin{figure}[!h]
%   \includegraphics[scale=0.2]{Figures/satisfaction.png}
%    \caption{Mean satisfaction level for each facility with respect to neighbourhood}
%   \label{fig:msl_wrt_quest}
% \end{figure}

% \begin{figure}[!h]
% \centering
%   \includegraphics[scale=0.2]{Figures/top25.png}
%    \caption{Facilities wise and overall relative quality improvement while moving from one neighbourhood to another } 
%   \label{fig:rqi}
% \end{figure}
	
  \label{fig:migration_stats}
  
\begin{figure}[!htb]
  \centering
  \includegraphics[scale=0.13]{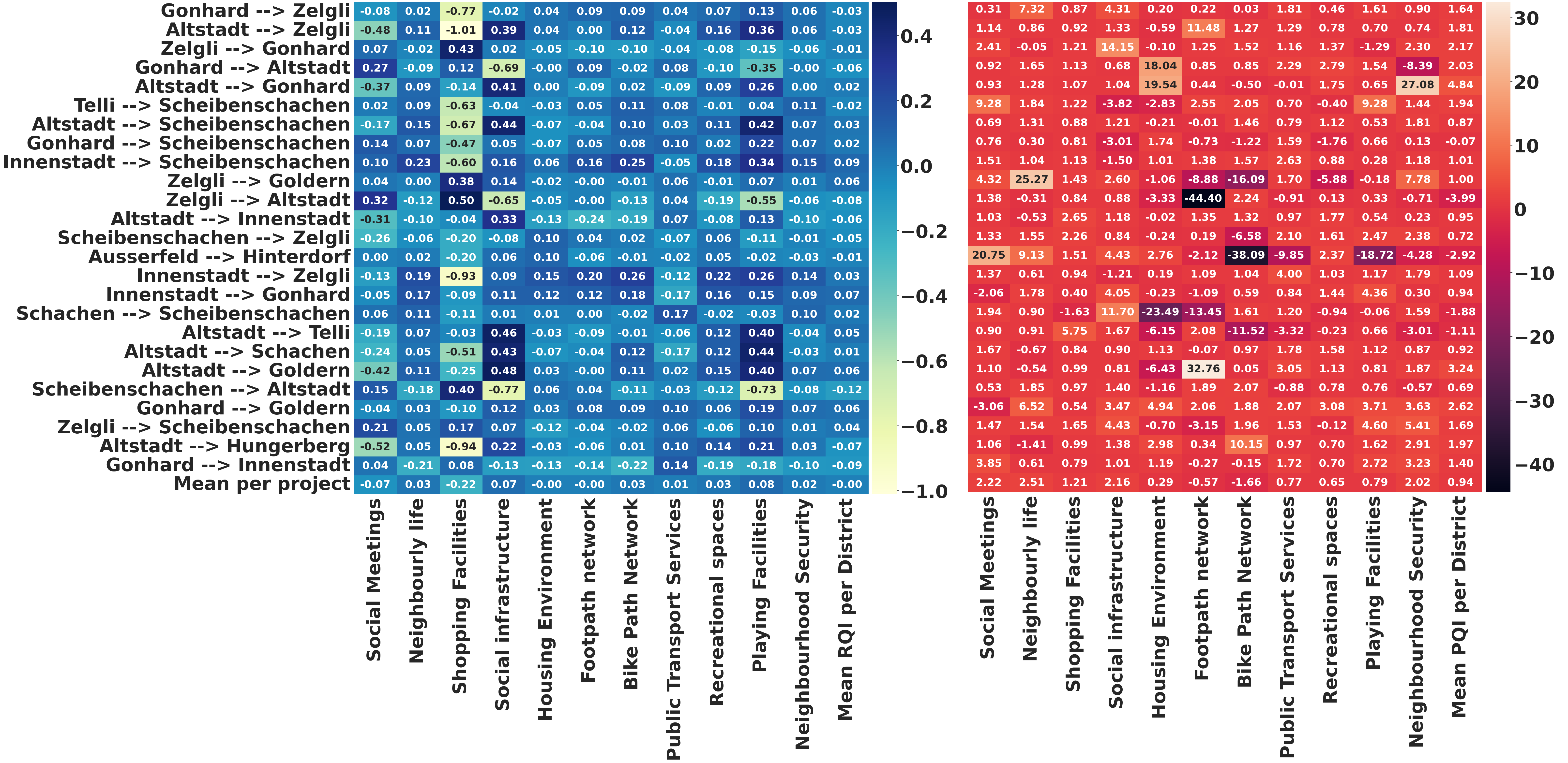}
  \captionsetup{justification=centering}
  \caption{Relative quality improvement (left) and perceived quality improvement (right) of project sectors by relocation.}
  \label{fig:rqi_pqi}
\end{figure}

 The order of relocations within Figure~\ref{fig:rqi_pqi} is ranked by their likelihood of occurrence. For example, Fig.~\ref{fig:migration_stats1} outlines the most frequent relocation is from \textit{Gonhard} to \textit{Zelgli}, which has a normalized value of 1.00. We can see that the greatest improvement in overall $RQI$ comes with relocations from \textit{Innenstadt} to \textit{Scheibenschachen}. When we look, specifically, at the scores for each project sector, we can see there is a significant decrease in the $RQI$ score for shopping facilities (-0.6) and a small decrease in public transport services (-0.05), which corresponds with the type of facilities present within these neighborhoods~\cite{quartierentwicklungskonzept_2020}. However, there is an increase in $RQI$ in all other metrics, with the most significant increase coming in playing facilities (0.34), which seems to correspond with an increased number of playing facilities within the neighborhood of \textit{Scheibenschachen}~\cite{quartierentwicklungskonzept_2020}. The most significant deterioration in $RQI$ is found in relocations from \textit{Scheibenschachen} to \textit{Altstadt} (which is the centre of Aarau). In some sense, this is relatively surprising, as \textit{Altstadt} can be seen as having the most comprehensive level of public facilities~\cite{quartierentwicklungskonzept_2020}. There is an increase in shopping facilities (0.4), however, there is also a significant decrease in playing facilities (-0.73), which corresponds to the type of facilities in both \textit{Scheibenschachen} and \textit{Altstadt} \cite{quartierentwicklungskonzept_2020}.

Moving further, we calculated the mean $RQI$ for each project sector and every relocation to rate how project sectors improve and deteriorate in all these relocations. Eq.~\ref{eq:rqi} yields positive $RQI$ score for \textit{Playing Facilities, Social Infrastructure, Recreational Spaces, Niehgbourly Life and Public Transport Services}. However, with the improvement in these facilities, people compromise on other facilities, which include \textit{Housing Environment, Social Meetings and Shopping Facilities}. In some sense, this supports our finding that shopping facilities do not significantly impact citizens' quality of life. %However, it is unclear as to why citizens are willing to compromise on other sectors.  

The $PQI$ measurement in Figure~\ref{fig:rqi_pqi} assesses the perceived improvement or deterioration in the quality of life for each individual when relocating from one neighbourhood to another, which takes into account citizens' satisfaction levels with various facilities and services. The $PQI$ scores is more positive than the $RQI$ scores. This suggests, that on average, citizens perceive a greater improvement in their quality of life when relocating from one neighborhood to another as compared to the actual improvement in the $RQI$. Interestingly, relocation from \textit{Zelgli} to \textit{Altstadt} comes with a significant reduction in the perceived quality of footpath networks. When these findings are considered in relation to Fig.~\ref{fig:distsan}, it can be observed that the greatest legitimacy gain in \textit{Altstadt} comes from project types that corresponds with footpath networks (such as public transport and parks).

\section{Conclusion and Future Work}\label{sec:Conclusion}

Policymakers' capacity to be responsive to citizens requires an ability to maintain legitimacy, whilst also improving the quality of life and satisfaction for citizens. In order to provide guidance to city authorities, in this paper, we propose a legitimacy measure to determine the feasibility of participatory interventions. In doing so, we demonstrate the optimal number of project sectors to fund that maximize legitimacy. Specifically, this has allowed us to demonstrate how the neighborhoods of \textit{Altstadt, Scheibenschachen, Goldern, Binzenhof} and \textit{Ausserfeld} may be considered as suitable candidates for participatory intervention - on the basis that the local authority is less-likely to meet the minimum funding requirements for these neighborhoods. In addition, we are able to identify the degree of legitimacy should the local authority invest in project types across a number of neighborhoods. We then highlight the most important project sectors for satisfaction and participation in the neighbourhoods that explain the citizens' overall quality of life. Using a data-driven approach, our research contributes to an understanding of the type of intervention that may benefit citizens' quality of life. We find that shopping facilities and public transport are unlikely to impact citizens quality of life, which may offer guidance for policymakers when considering funding allocation. Finally, through the exploration of relocation data, we are able to identify the importance citizens place on particular project sectors and link satisfaction to the likelihood for citizens to relocate. We find that citizens are most likely to compromise on the quality of shopping facilities, which supports the findings of the  classification model. We also find that, in general, citizens' perception of improvement in facilities when relocating is positive. 

The results outlined within this paper can open up new directions for future research. As previously outlined, our research informs the design of an upcoming novel participatory budgeting campaign in Aarau, Switzerland. Future research includes the study of additional factors that may impact satisfaction and legitimacy, such as local economic development, historical voting patterns, polling data etc. Intelligent pervasive data collection and decision-support systems for participation and collective decision-making can assist policy-makers to preserve legitimacy on a continuous basis, as we aspire to achieve with ongoing research on this area~\cite{Pournaras2020,Pournaras2021,Castells2020,Helbing2022}. Finally, further joint data analysis is required to understand, in more depth, the motivation behind intra-city relocations and how they relate to satisfaction or gentrification policies. 

Finally, it should be considered that citizens' input may not always be the most reliable source of guidance for policymakers. For example, citizens may be impacted by media consumption \cite{hoewe2020power} and anecdotal information \cite{moore2009experts}. While policymakers need to reflect the wishes of the citizenry, they also need to rely on expert advice to ensure accurate decision making is undertaken \cite{christensen2021expert}. Therefore, there is the potential for a limited degree of convergence between citizens' wishes and policymakers' decisions, even when policymakers are informed of citizens desires. With that in mind, policymakers may wish to utilize the approach of this paper, but may also take additional steps to enhance legitimacy, considering their approach to public relations, information sharing, and balancing public demand with expert advice. 

\section*{Acknowledgements}\label{sec:Acknowledgment}

We would like to thank Neus Llop Torrent and Faraz Awan for preliminary insights on this project. The authors would also like to thank Selina Frey, Lea Schneidegger and Jasmin Odermatt for providing access and insights on the data. This work is supported by a UKRI Future Leaders Fellowship (MR\-/W009560\-/1): `\emph{Digitally Assisted Collective Governance of Smart City Commons--ARTIO}' and the SNF NRP77 `Digital Transformation' project "Digital Democracy: Innovations in Decision-making Processes", \#407740\_187249. 

\bibliographystyle{ACM-Reference-Format}
\bibliography{Ref}

\end{document}